\documentclass[12pt]{article}
\usepackage[dvips]{graphics}
\usepackage{graphicx}
\usepackage{amsmath}
\usepackage[body={5.9in, 9.0in},left=1.350in, top = 1.35in]{geometry}
\usepackage{setspace}
\usepackage{cite}
\usepackage{amssymb}
\newcommand{\x}{\textit{x} }

\newcommand{\p}{\partial}
\newcommand{\be}{\begin{equation}}
\newcommand{\ee}{\end{equation}}
\newcommand{\bea}{\begin{eqnarray}}
\newcommand{\eea}{\end{eqnarray}}
\newcommand{\bes}{\begin{equation*}}
\newcommand{\ees}{\end{equation*}}
\newcommand{\ba}{\begin{array}}
\newcommand{\ea}{\end{array}}

\begin{document}

\title{ Comparative analysis of analytical solutions for $F_2^P(x,t)$ at small \textit{x} in DGLAP approach}
\author{D. K. Choudhury$^{a,b}$, Neelakshi N. K. Borah$^a$,  \\
$^a$ Department of Physics, Gauhati University, Guwahati-781014, India\\
$^b$ Centre for Theoretical Physics, Pandu College, Guwahati, India\\
E-mail:  $nishi_-indr@yahoo.co.in$}
\date{}
\maketitle
\begin{abstract}

Coupled DGLAP equations involving singlet quark and gluon distributions are explored by a Taylor expansion at small \textit{x} as two first order partial differential equations in two variables : Bjorken \textit{x} and \textit{t} ($t =ln\frac{Q^2}{\Lambda^2}$). The system of equations are then solved by the Lagrange's method and Method of Characteristics. We obtain the proton structure function $F_2^P(x,t)$ by combining the
corresponding non-singlet and singlet structure functions by both the methods.
Analytical solutions for $F_2^P(x,t)$ thus obtained are compared with the recent data published by H1 and ZEUS as well as with NNPDF3.0 parametrization and their compatibility is checked. Comparative analysis favors the analytical solution by Lagrange's method.\\
Keywords:Deep inelastic scattering, DGLAP equations, pQCD.\\
PACS Nos: 12.38.-t;12.38.B\_x;13.60.-r;13.60.Hb

\end{abstract}
\section{Introduction}
\label{ch5intro}
In our earlier work \cite{Borah:2013hoa}, we have made an extensive comparative study on the applicability of the two analytical methods: Lagrange's method and method of characteristics in context of the unpolarised non-singlet structure function $F_2^{NS}$. We have discussed different kinematic regions in which both the methods showed validity by comparing with both data and exact results for non-singlet sector. These analytical methods can be extended simply to embrace the singlet sector, so that it can be used to find the gluon and quark distribution functions. Solutions of DGLAP \cite{Dokshitzer:1977sg,Gribov:1972ri,Lipatov:1974qm,Altarelli:1977zs} evolution equations give quark and gluon structure functions which produce ultimately proton, neutron and deuteron structure functions. The standard program to study the $x$ dependence of quark and gluon PDFs is to compare the numerical solutions of the DGLAP equations with the data and so to fit the parameters of the $x$ profiles of the PDFs at some initial scale $Q_0^2$ and the asymptotic scale parameter $\Lambda$. However, for analyzing exclusively the small-$x$ region, there exists alternative simpler analysis, yielding analytical solutions of the DGLAP equations \cite{Ball:1994du,Ball:1994kc,Kotikov:1998qt,Mankiewicz:1996sd,Illarionov:2004nw}. Some approximated analytical solutions of DGLAP evolution equations suitable at small-$x$, have been reported in recent years \cite{Boroun:2005qx,Ghahramany:2002dn} with considerable phenomenological success.\\ 
Following the procedure of ref.\cite{Borah:2013hoa}, in this present communication we have solved the integrodifferential equation for the quark and gluon distribution functions in the leading order (LO), using the analytical methods and construct our analytical solutions for proton structure function $F_2^P(x,t)$ as the sum of a flavor nonsinglet $F_2^{NS}$ and a flavor singlet $F_2^{S}$ distribution.
In the experimental front, H1 \cite{Andreev:2013vha} and ZEUS \cite{Abramowicz:2014jak} collaboration has published very recent data on $F_2^P(x,t)$ covering wide kinematic region of Bjorken $x$, and low to medium four-momentum transfer squared, $Q^2$. In H1 \cite{Andreev:2013vha} collaboration data, the earlier measurements of $F_2^P(x,t)$ are superseded by the recently published data in the kinematic region $6.5 \times 10^{-4}\leq x \leq 0.32 \times 10^{-1}$ and $1.5GeV^2 \leq Q^2 \leq 800GeV^2$, whereas in ZEUS \cite{Abramowicz:2014jak} $F_2^P(x,t)$ is measured in the region $0.00025\leq x \leq 0.00493$ with $9 GeV^2 \leq Q^2 \leq 110 GeV^2$. In this paper we have discussed the relative compatibility of both the analytical methods in context of $F_2^P(x,t)$ at leading order (LO) with the recent data from H1 and ZEUS experiment as well as with the latest NNPDF3.0 \cite{Ball:2014uwa} parametrization based on HERA Run-II data. A newer data-set allows the extension of the kinematic range towards lower values of $x$ for comparative analysis, in which our analytical models are approximated at. \\
In section 2 we describe the formalism, section 3 is devoted to testing our prediction's comparison with the  data, while in section 4, we give our conclusion.

\section{Formalism \label{formalism}}
\subsection{Singlet coupled DGLAP equations in Taylor approximated form}The coupled DGLAP equations  for quark singlet $(\Sigma(x,Q^2))$ and gluon $(G(x,Q^2))$ densities are \cite{Dokshitzer:1977sg,Gribov:1972ri,Lipatov:1974qm,Altarelli:1977zs},
\be
\label{eqn:singcoup}
\frac{\p}{\p \ln Q^2}\left(\ba{c}
\Sigma\left(x,Q^2\right)\\
G\left(x,Q^2\right)
\ea
\right) = \frac{\alpha_s\left(Q^2\right)}{2\pi}\left(
\ba{cc}
P_{qq} & P_{qg} \\
P_{gq} & P_{gg}
\ea
\right)\otimes \left(
\ba{c}
\Sigma\left(x,Q^2\right) \\
G\left(x,Q^2\right)
\ea
\right) ,
\ee
where $\alpha_s\left(Q^2\right)$ is the strong coupling constant, $ P_{i,j}$s are the Altarelli-Parisi splitting functions and the symbol $\otimes$ stands for the usual Mellin convolution in the first variable defined as,
\be
\label{eqn:ch5Mellin}
 a(x)\otimes f(x)=\int_x^1\,\frac{dy}{y}a(y)f\left(\frac{x}{y}\right).
\ee
Introducing the variable $ t=\ln\frac{Q^2}{\Lambda^2}$ and using the explicit forms of the splitting functions  $ P_{i,j}(i,j=q,g)$ in LO, the evolution equation for singlet distribution can be written as \cite{Abbott:1979yb},
\begin{eqnarray}
\label{eqn:ch5F2s1}
&\displaystyle{\frac{\p F_2^S(x,t)}{\p t}}-\frac{A_f}{t}\left[\{3+4\ln(1-x)\}F_2^S(x,t)+2\int_x^1\frac{dz}{(1-z)}\left\{(1+z^2)F_2^S\left(\frac{x}{z},t\right)\right.\right. \nonumber \\
& \left.\left.-2F_2^S(x,t)\right\}+\frac{3}{2} n_f \int_x^1dz\left(z^2+(1-z)^2\right)G\left(\frac{x}{z},t\right) \right] =  0
\end{eqnarray}
Similarly the DGLAP evolution equation for non-singlet distribution can be written as,
\begin{eqnarray} 
\label{eqn:AP}
\frac{\p F_2^{NS}(x,t)}{\p t}-\frac{A_f}{t}\left[\left\{3+4\ln(1-x)\right\}F_2^{NS}(x,t)\right. & & \nonumber \\
 \left.
 +2\int_x^1\frac{dz}{1-z}\left\{(1+z^2)F_2^{NS}\left(\frac{x}{z},t\right)-2F_2^{NS}(x,t)\right\}\right]=0.
\end{eqnarray} 
Here $A_f=\frac{4}{3\beta_0}$, $ \beta_0=11-\frac{2}{3}n_f $, $n_f$ being the number of flavors considered and  
$\alpha_s(t)=\frac{4\pi}{\beta_0 t}$. $F_2^S(x,t)$ and $F_2^{NS}(x,t)$ are the singlet and non-singlet structure functions of the proton.
We write Eq.(\ref{eqn:ch5F2s1}) and Eq.(\ref{eqn:AP}) as,
\begin{eqnarray}
\label{eqn:ch5F2s2}
\displaystyle{\frac{\p F_2^S(x,t)}{\p t}}-\frac{A_f}{t}\left[\{3+4\ln(1-x)\}F_2^S(x,t)+I_1^S(x,t)+I_1^G(x,t)\right]=0
\end{eqnarray}
\begin{equation}
\label{eqn:AP1}
\frac{\p F_2^{NS}(x,t)}{\p t}=\frac{A_f}{t}\left[\left\{3+4\ln(1-x)\right\}F_2^{NS}(x,t) +I_1^{NS}(x,t)\right]
\end{equation}
where
\begin{equation}
\label{ch5I1s}
I_1^S(x,t)=2\int_x^1\frac{dz}{1-z}\left[(1+z^2)F_2^S(\frac{x}{z},t)-2F_2^S(x,t)\right] ,
\end{equation}
\be
\label{eqn:ch5I1g}
I_1^G(x,t)=\frac{3}{2} n_f \int_x^1 dz\,[z^2+(1-z)^2]G(\frac{x}{z},t) ,
\ee
\be
\label{Ch5I1NS}
I_1^{NS}=2\int_x^1\frac{dz}{1-z}\left\{(1+z^2)F_2^{NS}\left(\frac{x}{z},t\right)-2F_2^{NS}(x,t)\right\}
\ee
To carry out the integrations in Eqs.(\ref{ch5I1s}-\ref{Ch5I1NS}), we introduce the variable $ u$  defined as $u=1-z $ and expand the argument $\displaystyle{\frac{x}{z}}$ as a series.
\be
\label{eqn:ch5xbyz}
\frac{x}{z}=\frac{x}{1-u}=x \sum_{k=0}^\infty u^k=x+x\sum_{k=1}^{\infty} u^k 
\ee
Since $x<z<1$, so $0<u<1-x$ ; hence the series is convergent for $\vert u\vert<1$ and we can use Taylor expansion of $F_2^S(\frac{x}{z},t)$, $G(\frac{x}{z},t)$ and $F_2^{NS}(\frac{x}{z},t)$ in a approximated form and as $x$ is small in our region of discussion, the terms containing $x^2$ and higher powers of $x$ can be neglected as those terms are still smaller and
therefore, we can rewrite,
\be
\label{eqn:ch5F2series}
F_2^{S,NS}(\frac{x}{z},t)\approx F_2^{S,NS}(x,t)+x\sum_{k=1}^\infty u^k \frac{\p F_2^{S,NS}(x,t)}{\p x}
\ee
\begin{equation}
\label{eqn:ch5F2seriesG}
G(\frac{x}{z},t)\approx G(x,t)+x\sum_{k=1}^\infty u^k \frac{\partial G(x,t)}{\partial x}
\end{equation}
where the terms containing $x^2$ and higher powers of $x$ are neglected at small $x$.\\
Using the above Eq.(\ref{eqn:ch5F2series}) and Eq.(\ref{eqn:ch5F2seriesG}) we carry out the integrations in $z$ in Eq.(\ref{ch5I1s}-\ref{Ch5I1NS}). Neglecting terms $ \mathcal{O}(x^2)$ which is justified at small \x, we get,
\be
\label{eqn:ch5I1sAp}
I_1^S(x,t)\approx(2x-3)F_2^S(x,t)+\left(x+2 x \ln\frac{1}{x}\right)\frac{\p F_2^S(x,t)}{\p x},
\ee
\be
\label{eqn:ch5I1gAp}
I_1^G(x,t)\approx n_f\left(1-\frac{3}{2}x\right)G(x,t)-\frac{n_f}{2}\left(5x-3 x \ln\frac{1}{x}\right)\frac{\p G(x,t)}{\p x},
\ee
\be
\label{eqn:ch5I1NSAp}
I_1^{NS}(x,t)\approx \left[2\ln(\frac{1}{x})+(1-x^2)\right]-\left(x-1 \right)\left(x+3 \right)F_2^{NS}(x,t) ,
\ee
The exact relation between the gluon distribution  $G(x,t)=xg(x,t)$ and quark distribution $F_2^S(x,t)=x\sum_i e_i^2\left\lbrace q_i(x,t)+\bar{q_i}(x,t)\right\rbrace $ is not derivable in QCD even in LO. However, simple forms of such relation are available in literature to facilitate the analytical solution of coupled DGLAP equations. In ref \cite{Choudhury:1989sj}, it was assumed that $Q^2$ dependence of both the distributions are identical.
In ref. \cite{Sarma:1997du}, on the other hand, the following simple relation was assumed,
\begin{equation}
\label{approx1}
G(x,Q^2)=k.F_2^S(x,Q^2)
\end{equation}
where parameter $k$ has to be determined from experiments.\\
Again as the input singlet and gluon parametrization, taken from global analysis of parton distribution functions, which incorporate different high precision data, are also functions of $x$ at fixed $Q^2$, hence 
relation between singlet structure function and gluon parton densities may be expressed as a function of $x$ \cite{Devee:2012zz}.\\
However a more rigorous analysis was done by Lopez and Yndurain \cite{Lopez:1979bb} and they investigated the behavior of the singlet $F_2^S(x,Q^2)$ and gluon $G(x,Q^2)$ as $x\rightarrow 0$. They observed that 
\begin{equation}
F_2^S(x,Q^2)_{x\rightarrow 0}=B_S(Q^2)x^{-\lambda_S}
\end{equation}
\begin{equation}
G(x,Q^2)_{x\rightarrow 0}=B_G(Q^2)x^{-\lambda_G}
\end{equation}
where $B_S$ and $B_G$ are $Q^2$ dependent as $\lambda _G = \lambda _S$ and $\lambda _S$ is strictly positive. Thus,
\begin{equation}
\frac{G(x,Q^2)}{F(x,Q^2)}_{x\rightarrow 0}\simeq f(Q^2)
\end{equation}
It suggests a more general form \cite{0253-6102-61-5-20} following,
\begin{equation}
\label{rel1}
G(x,Q^2)=K(Q^2)F_2^S(x,Q^2)
\end{equation}
than Eq.(\ref{approx1}).\\
Now using Eq.(\ref{eqn:ch5I1sAp}) and Eq.(\ref{eqn:ch5I1gAp}) we can express Eq.(\ref{eqn:ch5F2s1}) in a more precise form as,
\begin{eqnarray}
\label{singletpde}
\frac{\p F_2^S(x,t)}{\p t}-\frac{A_f}{t}\left[{3+4\ln(1-x)}F_2^S(x,t)+(2x-3)F_2^S(x,t)+\left(x+2 x \ln\frac{1}{x}\right)\frac{\p F_2^S(x,t)}{\p x}\right] & & \nonumber\\
 -\frac{A_f}{t}\left[ n_f\left(1-\frac{3}{2}x\right)G(x,t)-\frac{n_f}{2}\left(5x-3 x \ln\frac{1}{x}\right)\frac{\p G(x,t)}{\p x}\right]=0
\end{eqnarray}
Similarly using Eq.(\ref{eqn:ch5I1NSAp}) we write for the non-singlet $F_2^{NS}$ as,
\begin{eqnarray}
\label{firstapprox}
\frac{\p F_2^{NS}(x,t)}{\p t}-\frac{A_fx}{t}\frac{\p F_2^{NS}(x,t)}{\p x}\left[2\ln(\frac{1}{x})+(1-x^2)\right]-\frac{A_f}{t}\left[{3+4\ln(1-x)}\right. & & \nonumber\\
 \left.
+\left(x-1 \right)\left(x+3 \right)\right]F_2^{NS}(x,t)=0
\end{eqnarray} 
Eq.(\ref{firstapprox}) is a partial differential equation for the non-singlet structure function $F_2^{NS}(x,t)$ with respect to the variables $x$ and $t$. The solutions of the Eq.(\ref{firstapprox}) by both Lagrange's and Method of Characteristics has been reported in our earlier work \cite{Borah:2013hoa}. So we don't discuss the solution of the non-singlet structure function $F_2^{NS}(x,t)$ any further in this work. We continue our discussion to obtain the analytical solutions for Eq.(\ref{singletpde}).
Using above relation given by Eq.(\ref{rel1}) we express Eq.(\ref{singletpde}) as,
\begin{eqnarray}
\label{singletpde1}
\frac{\p F_2^S(x,t)}{\p t}-\frac{A_f}{t}\left[\left\lbrace 3+4\ln(1-x)+(2x-3)\right\rbrace F_2^S(x,t)+n_f\left(1-\frac{3}{2}x\right)K(Q^2)F_2^S(x,t)\right]  & & \nonumber\\
-\frac{A_f}{t}\left[x+2x \ln \frac{1}{x}-\frac{n_f}{2}\left(5x-3 x \ln\frac{1}{x}\right) K(Q^2) \right]\frac{\p F_2^S(x,t)}{\p x}=0
\end{eqnarray}
which is a partial differential equation for the singlet structure function $F_2^S(x,t)$ with respect to the variables  $x$ and $t$. We solve this PDE Eq.(\ref{singletpde1}) with the two formalisms described here, the Lagrange's method and Method of Characteristics. In order to do that we express Eq.(\ref{singletpde1}) as,
\begin{eqnarray}
\label{singletpde2}
t\frac{\p F_2^S(x,t)}{\p t}=\omega_1\frac{\p F_2^S(x,t)}{\p x}+\omega_2 F_2^S
\end{eqnarray}
where
\begin{equation}
\omega_1=\frac{4}{3\beta_0}\lbrace x+2x\ln \frac{1}{x}-\frac{n_f}{2}\left(5x-3 x \ln\frac{1}{x}\right) K(Q^2)\rbrace
\end{equation}
\begin{equation}
\label{omega2ch3}
\omega_2=\frac{4}{3\beta_0}\lbrace 3+4\ln(1-x)+(2x-3)+n_f\left(1-\frac{3}{2}x\right)K(Q^2)\rbrace
\end{equation}
\subsection{Solution by the Lagrange's Auxiliary Method}
To solve the equation Eq.(\ref{singletpde2}) by the Lagrange's Auxiliary method \cite{sneddon2006elements}, we write the equation in the form,
\be
\label{eqn:lagauxeqn}
Q(x,t)\frac{\p F_2^{S}(x,t)}{\p t}+P(x,t)\frac{\p F_2^{S}(x,t)}{\p x}=R(x,t,F_2^{S}(x,t))
\ee
where
\be
Q(x,t)=t
\ee
\be
P(x,t)=-\omega_1
\ee
and
\be
R(x,t,F_2^{S}(x,t))=R'(x)F_2^{S}(x,t)
\ee
with
\be
R'(x)=\omega_2
\ee
The general solution of the Eq.(\ref{eqn:lagauxeqn}) is obtained by solving the following auxiliary system of ordinary differential equations,
\begin{equation}
\label{pde1}
\frac{dx}{P(x)}=\frac{dt}{Q(t)}=\frac{dF_2^{S}(x,t)}{R(x,t,F_2^{S}(x,t))}
\end{equation}
If $u(x,t,F_2^{S})=C_1$ and $v(x,t,F_2^{S})=C_2$ are the two independent solutions of Eq.(\ref{eqn:lagauxeqn}), then in general, the solution of Eq.(\ref{eqn:lagauxeqn}) is
\begin{equation}
F(u,v)=0
\end{equation}
where $F$ is an arbitrary function of $u$ and $v$.\\
 In this approach we try to find a specific solution that satisfies some physical conditions on the structure function. Such a solution can be extracted from the combination of $u$ and $v$ linear in $F_2^{S}$, the simplest possibility being,
 \begin{equation}
 \label{linear}
 u+\alpha v=\beta
 \end{equation}
 where $\alpha$ and $\beta$ are two quantities to be determined from the boundary conditions on $F_2^{S}$. Solving Eq.(\ref{pde1}), we obtain,
\begin{equation}
u(x,t,F_2^{S})=tX^{S}(x)
\end{equation}
and
\begin{equation}
v(x,t,F_2^{S})=F_2^{S}(x,t)Y^{S}(x)
\end{equation}
The functions $X^{S}(x) $ and $Y^{S}(x)$ are defined as
\begin{equation}
\label{Xs1}
X^{S}(x)=\exp[-\int \frac{dx}{P(x,t)}]
\end{equation}
\begin{equation}
Y^{S}(x)=\exp [-\int \frac{R'(x)}{P(x,t)}dx]
\end{equation}
The explicit analytical form of $X^{S}(x)$ in the leading $(\frac{1}{x})$ approximation (at very small $x$ region $\log(\frac{1}{x})\gg x\log(\frac{1}{x})\gg x $) comes out to be,
\begin{equation}
\label{Xnsvalue1}
X^{S}(x)=\exp[\frac{6\beta_0}{4(4+3n_fK(Q^2)}\log [\log x]]
\end{equation}
 Using the physically plausible boundary conditions for structure functions,i.e.
\begin{eqnarray}
F_2^S(x,t)&=&F_2^S(x,t_0), \text{for} \quad t=t_0\\
F_2^S(1,t)&=&0 \quad \text{for any} \quad t.
\end{eqnarray} 
and putting the values of $u$ and $v$ in Eq.(\ref{linear}), we obtain the solution for Eq.(\ref{eqn:lagauxeqn}) as,
\be
\label{eqn:solnlagmethod}
F_2^{S}(x,t)=F_2^{S}(x,t_0)\left(\frac{t}{t_0}\right)\frac{[X^{S}(x)-X^{S}(1)]}{[X^{S}(x)-(\frac{t}{t_0})X^{S}(1)]}
\ee 
Eq.(\ref{Xnsvalue1}) gives us,
\begin{equation}
X^{S}(1)=0
\end{equation}
which yields,
\begin{equation}
\label{solnsinglet1}
F_2^{S}(x,t)=F_2^{S}(x,t_0)\left(\frac{t}{t_0}\right)
\end{equation}
 Eq.(\ref{solnsinglet1}) gives the $t$ evolutions of singlet structure
function at LO and is the solution for $F_2^S$ by Lagrange's method.
\subsection{Solution by the method of characteristics}
To solve the PDE Eq.(\ref{singletpde2}) by the method of characteristics \cite{farlow2012partial,williams1980partial}, we express it in terms of a new set of coordinates $(s,\tau)$, such that Eq.(\ref{singletpde2}) becomes an ODE w.r.t. one of the new variables.
We know that most of the important properties of the solution of Eq.(\ref{singletpde2}) depends on the principal part of the equation i.e. the left hand side in Eq.(\ref{eqn:lagauxeqn}). This part is actually a total derivative along the solution of the characteristic equation, 
\begin{equation}
\label{characterics1}
\frac{dx}{dt}=-\frac{\omega_1}{t}
\end{equation}
which gives the characteristic curves of Eq.(\ref{singletpde2}). That is along the characteristic curve, the partial differential equation becomes an ordinary differential equation.\\
 The characteristic equation Eq.(\ref{characterics1}) can be written as,
 \begin{equation}
 \frac{dx}{dt}=\frac{dx}{ds}\frac{ds}{dt}
 \end{equation}
 with,
\begin{equation}
\frac{dt}{ds}=t
\end{equation},
\begin{equation}
\frac{dx}{ds}=-\omega_1
\end{equation}
Using Eq.(\ref{characterics1}) in Eq.(\ref{singletpde2}), the left hand side becomes an ordinary derivative with respect to $s$ and the equation becomes an ordinary differential equation,
\begin{equation}
\label{charactcurve}
\frac{dF_2^{S}(s,\tau)}{ds}+c^{S}\left(s,\tau \right)F_2^{S}(s,\tau)=0
\end{equation}
where\\
\begin{equation}
c^{S}(s,\tau)=\omega_2
\end{equation}
$\omega_2$ is as given above in Eq.(\ref{omega2ch3}).\\
As per the initial conditions i.e. $x(s=0)=\tau$ and $t(s=0)=t_0$, the solutions of the characteristics equations yield,
\begin{eqnarray}
\label{solucharac}
s=\log\log(\frac{\tau}{x})^{\alpha_1} \nonumber\\
\tau=x\exp\left[(\frac{t}{t_0})^{\frac{1}{\alpha_1}}\right] 
\end{eqnarray}
with,
\begin{equation}
\alpha_1=\frac{3\beta_0}{4\lbrace 2+K(Q^2)\frac{9}{2}\rbrace}
\end{equation}
Integrating Eq.(\ref{charactcurve}) along the characteristic curve, we obtain the solution for $F_2^{S}(x,t)$ in $(s,\tau)$ space as,
\begin{equation}
\label{mocsinglet1}
F_2^S(s,\tau)=F_2^S(\tau)(\frac{t}{t_0})^{-\frac{4}{3\beta_0}( \xi_1)}
\end{equation}
where,
\begin{equation}
\label{exponent1}
\xi_1 =4\log\left( 1-\tau\exp\left[-(\frac{t}{t_0})^{\frac{1}{\alpha_1}}\right] \right)+2\tau\exp\left[-(\frac{t}{t_0})^{\frac{1}{\alpha_1}}\right]
+n_f\left( 1-\frac{3}{2}\tau\exp\left[-(\frac{t}{t_0})^{\frac{1}{\alpha_1}}\right]\right) K(Q^2)
\end{equation}
Using the solutions of the characteristic equations Eq.(\ref{solucharac}), which lead us to the $(x,t)$ space, we can express Eq.(\ref{mocsinglet1}) in a more precise form as,
\begin{equation}
\label{mocsinglet}
F_2^S(x,t)=F_2^S(x,t_0)(\frac{t}{t_0})^{n(x,t)}
\end{equation}
where,
\begin{equation}
n(x,t)=-\frac{4}{3\beta_0}( \xi_1)
\end{equation}
and $F_2^S(\tau)=F_2^S(x,t_0)$ is the input function.
Eq.(\ref{mocsinglet}) is the analytical solution for the singlet
structure function within the present formalism. Unlike Eq.(\ref{solnsinglet1}), Eq.(\ref{mocsinglet}) is sensitive to gluon distribution as well as $K(Q^2)$, which occurs in spite of the leading log approximations (at very small $x$ region $(\log(\frac{1}{x})\gg x\log(\frac{1}{x})\gg x)$). The reason is that in the Lagrange's method, $F_2^{NS}$ \cite{Borah:2013hoa} and  $F_2^{S}$ has identical evolution for the least approximated level. It is possible only when the gluon effect is absent. But this feature has already been well observed \cite{Sarma:1997du,Choudhury:1992nv}, the only new observation is that it is true even when $k=K(Q^2)$, i.e. $Q^2$ dependent.\\
Using our results derived in this section, we will calculate the proton structure function $ F_2^P(x,t)$ from the singlet and the non-singlet structure function using the relation
\be
\label{eqn:ch5F2p}
F_2^P=\frac{3}{18}F_2^{NS}+\frac{5}{18}F_2^S\
\ee
We discuss in the next section the phenomenological consequences of our results derived in this section. \\

\section{Form of $K(Q^2)$}
Let us now discuss the plausible forms of $K(Q^2)$ as defined in Eq.(\ref{rel1}) above and discuss the related constraints on the parameter duos $k$ and $\sigma$.
\subsection{Choice of the form of $K(Q^2)$}
The important characteristics of pQCD is the $\log Q^2$ dependence, as can be seen from the definition of running coupling constant, as well as any $Q^2$ evolution of structure function. However $Q^2$ alone does not appear and this basically yields the following plausible forms for the function $K(Q^2)$,
\begin{itemize}
\item[I.]
\begin{equation}
K(Q^2)=k\left( \frac{Q^2}{Q_0^2}\right)^\mu
\end{equation}
which is a power law in $Q^2$.
\item[II.]
\begin{equation}
K(Q^2)=k\left( \log\frac{Q^2}{\Lambda^2}\right)^\sigma 
\end{equation}
which gives us logarithmic dependence in $Q^2$.
\item[III.]
\begin{equation}
K(Q^2)=k\left( \log \log \frac{t}{t_0}\right)^\nu
\end{equation}
allowing us $\log \log Q^2$ dependence.
\end{itemize}
These forms are related to the other two possibilities through the correspondence,
\begin{equation}
\mu = \sigma \frac{\log t}{t} = \nu \frac{\log \log t}{\log t}
\end{equation}
While defining the forms of $K(Q^2)$, it is reasonable to have a growth of `$t$' instead of $Q^2$. Possible generalized forms of $K(Q^2)$ are,
\begin{equation}
K(Q^2)=k\sum_{i=1}^{\infty} C_i\left( \frac{Q^2}{Q_0^2}\right)^{\mu_i}
\end{equation}
\begin{equation}
\label{form1}
K(Q^2)=k\sum_{i=1}^{\infty} C_i(t)^{\sigma_i}
\end{equation}
\begin{equation}
K(Q^2)=k\sum_{i=1}^{\infty} C_i\left( \log \frac{t}{t_0}\right)^{\nu_i}
\end{equation}
We have chosen the function $K(Q^2)$ with logarithmic dependence in $Q^2$, i.e. Eq.(\ref{form1}) for definiteness and simplicity. But such a pattern comes with large number of parameters. However such proliferation of parameters makes the phenomenology uninteresting. Hence economy of parameters in terms of numbers justify that the most appropriate and QCD inspired functional form for the function $K(Q^2)$ has to be of the logarithmic form and we consider it to be,
\begin{eqnarray}
\label{finalrel1}
K(Q^2)=k(\log \frac{Q^2}{\Lambda^2})^\sigma = kt^\sigma
\end{eqnarray}
where $k$ and $\sigma$ are two parameters to be fixed.\\
\subsection{Reality constraint on the parameters $k$ and $\sigma$}
The essential condition for our analytical solution for $F_2^S$, obtained by method of characteristics, Eq.(\ref{mocsinglet}) to be real is that the exponent $n(x,t)$ has to be real. This imposes a reality condition on $\xi_1$ which quantitatively leads us to the conclusion that,
\begin{eqnarray}
\label{realitycond}
0<\tau\exp\left[-(\frac{t}{t_0})^{\frac{1}{\alpha_1}}\right]< 1
\end{eqnarray}
which also follows from the definition of $\tau$ as given by Eq.(\ref{solucharac}). The new variable $\tau$ is dependent on both $k$ and $\sigma$ as $\alpha_1$ is parametrized by them. The inversely proportionate feature of $\alpha_1$ on both $k$ and $\sigma$ makes the realization that $k$ and $\sigma$ cannot be too large. So for any value of $x$ and $Q^2$, the choice of our parameter duos $k$ and $\sigma$ is bounded by the above reality condition and cannot be treated as free parameters.\\
The recent HERA data allows us to explore a wide range of $Q^2$ evolution,i.e. $1.5 \leq Q^2 \leq 800 GeV^2$ for $F_2^P(x,t)$. Hence given the two parameters $k$ and $\sigma$, we fix them for this entire region of $Q^2$. The reality condition allows an effective range of values for both $k$ and $\sigma$ for the considered $Q^2$ region, the best fitted range for $k$ and $\sigma$ being $0.001< k< 1.45$ and $0.001< \sigma < 0.055$ respectively.\\
\section{Results and Discussion}
In this particular work we have calculated the $Q^2$ evolution for singlet structure functions using two analytical approaches. The proton structure function $F_2^P(x,t)$ has been calculated using the relation Eq.(\ref{eqn:ch5F2p}), extending our comparative study of the above discussed analytical methods in terms of proton structure function $F_2^P(x,t)$ with very recent experimental data published by H1 \cite{Andreev:2013vha} and ZEUS \cite{Abramowicz:2014jak} collaboration. We have used the LO MSTW 2008 \cite{Martin:2009iq} input for evolution of our solutions with $Q_0^2=1GeV^2$. We have plotted our analytical solution for $t$-evolution of $F_2^{P}(x,t)$ by Lagrange's method with H1 \cite{Andreev:2013vha} and ZEUS \cite{Abramowicz:2014jak} experimental data in figure 1 and figure 2 respectively. We have considered the range $10^{-4} \leq x \leq 10^{-2}$ and $2.5 GeV^2 \leq Q^2 \leq 500 GeV^2$ for H1 data and $0.00025 \leq x \leq 0.00493$ and $9 \leq Q^2 \leq 110 GeV^2$ for ZEUS data. As in case of ZEUS data, the extracted values of $F_2^{P}$ have been given only for 27 $(x,Q^2)$ bins against few fixed $Q^2$ values, hence we have considered the entire data range while plotting. Good consistency has been observed between our analytical solution by Lagrange's method and the experimental measurements by both H1 and ZEUS within the range $0.00025 \leq x \leq 0.001$ and $3 GeV^2 \leq Q^2 \leq 60 GeV^2$.\\
In figure 3 we have shown the $Q^2$ evolution of our analytical solution for $F_2^P(x,t)$ by method of characteristics, along with the H1 experimental data. As can be seen the evolution of the analytical solution by method of characteristics for $F_2^P(x,t)$ is not compatible with data, further we have observed that structure function has been decreasing with increasing $Q^2$. This behavior has been observed to be true for other regions of $x$ too as well as for ZEUS data, hence not included in the text here.\\
In figure 4 we have plotted the analytical models with the very recent NNPDF3.0 \cite{Ball:2014uwa} parametrization against $Q^2$ for different values of $x$. We have confined the comparison within the region $0.00025\leq x\leq 0.013$ and $3GeV^2\leq Q^2 \leq 85GeV^2$, for $F_2^P(x,t)$ produced by the NNPDF3.0 collaboration based on the HERA-II data. Here the vertical error bars represent uncertainties given by the standard deviations and computed by added in quadrature method in our work. We note that comparison with NNPDF3.0 parametrization also supports our observation with data and it is clear that the region of validity of our analytical solution by Lagrange's method is approximately in the range $0.0001 \leq x \leq 0.01$ and $3 GeV^2 \leq Q^2 \leq 60 GeV^2$. However our analytical solution by method of characteristics for $F_2^P(x,t)$ does not follow the general growth of evolution.\\
As per pQCD prediction structure function should rise in small $x$ with $Q^2$ \cite{Yndurain:1999ui,Ball:1994du,Ball:1994kc}. But from the form of ${\xi_1}$ as has been defined in Eq.(\ref{exponent1}), it is evident that for $x<0.66$ it is always positive and the exponent of our solution Eq.(\ref{mocsinglet}), i.e. $n(x,t)$ is negative. Thus the behavior of Eq.(\ref{mocsinglet}) contradicts both QCD expectation \cite{Yndurain:1999ui,Ball:1994du,Ball:1994kc} and data. Further if we extrapolate ${\xi_1}$ to large $x$ i.e. for $x \geq 0.66$, then it is observed that ${\xi_1}$ is negative yielding a positive exponent ${n(x,t)}$, which yields actually expectation of small $x$ QCD. The expected large $x$ behavior of QCD is that structure function should fall \cite{Yndurain:1999ui}. So the prediction extrapolated to high $x$ is not even consistent with QCD prediction. We therefore infer that method of characteristics is less favored than Lagrange's method as per the approximated solutions in DGLAP approach within the considered formalism, a feature which has presumably been overlooked earlier.\\
We note the sensitivity of the two parameters $k$ and $\sigma$ along with the Eq.(\ref{rel1}) towards $F_2^P(x,t)$, which have played a crucial role in case of method of characteristics. However in case of the solution by Lagrange's method, given by Eq.(\ref{solnsinglet1}), such restriction as $k$ and $\sigma$ does not appear in the solution. Effectively it means that Lagrange's method allows such growth with $Q^2$.\\
\section{Conclusion}
This work is an extension of the work of ref. \cite{Borah:2013hoa} for the two important analytical methods, Lagrange's and Method of Characteristics, in obtaining the analytical solutions for proton structure function $F_2^P(x,t)$, which consists of non-singlet $F_2^{NS}(x,t)$ and singlet $F_2^{S}(x,t)$ structure functions. For this part we pursue a general form as given by Eq.(\ref{rel1}), relating $F_2^{S}(x,Q^2)$ and $G(x,Q^2)$ for comparison with theoretical analysis of \cite{Lopez:1979bb}. However consequence of the relation could not be tested separately in the present work as it does not effect the analytical solution by the Lagrange's method, obtained in the leading $\log \frac{1}{x} $ approximation and only the solution by method of characteristics has exclusive dependence on the relation. We summarize our comparative analysis by looking at the impact of sensitive comparison with the three different sources of predictions for $F_2^P(x,t)$, H1, ZEUS and NNPDF3.0. However data analyzed in the range $0.00025 \leq x \leq 0.001$ and $3 GeV^2\leq Q^2 \leq 60 GeV^2$ is found to favor the former (Lagrange's method) and not the later (method of characteristics).\\
We conclude this section with a comment. We have followed the results of the general analysis of ref.\cite{Sarma:1997du} and used in \cite{Yndurain:1999ui} and not incorporated the observation of ref. \cite{Ball:2016spl}; i.e. for $x\rightarrow 0$ and $Q^2\rightarrow \infty$, the ratio $\frac{xG(x,Q^2)}{x(q+\bar{q})}\rightarrow \left[\frac{ln(\frac{1}{x})}{ln(\frac{t}{t_0})}\right]  ^{\frac{1}{2}}$, which suggests that the exponent $\sigma$ of Eq.(\ref{finalrel1}) might have additional $x$ and $Q^2$ dependence as well. Such effects are currently under study.
\begin{figure}[b]
\minipage{0.50\textwidth}
\includegraphics[width=\linewidth]{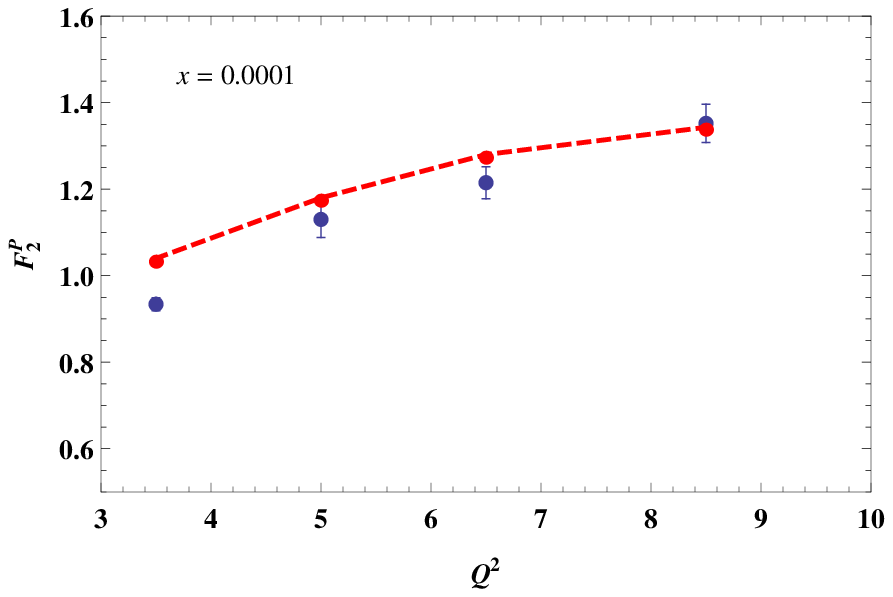}
\endminipage\hfill
\minipage{0.50\textwidth}
\includegraphics[width=\linewidth]{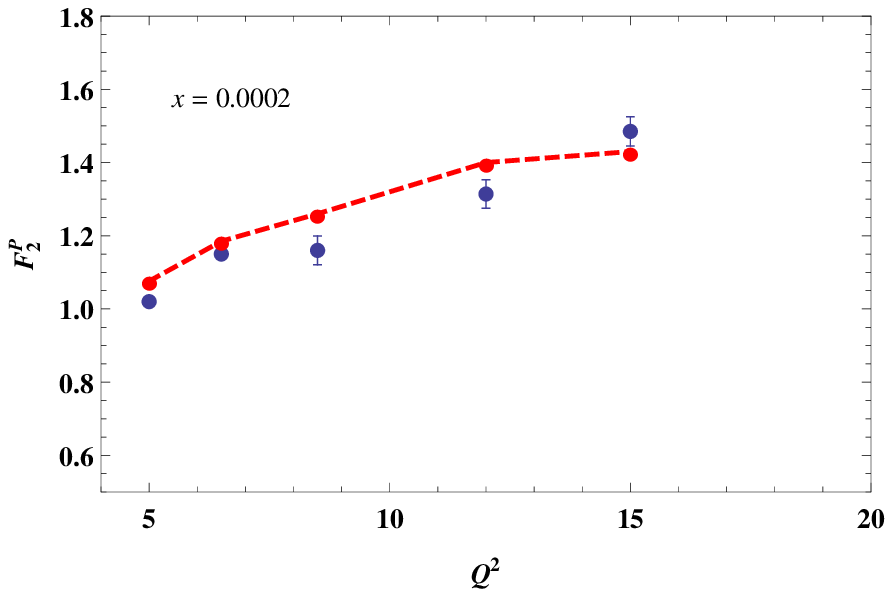}
\endminipage\hfill
\minipage{0.50\textwidth}
\includegraphics[width=\linewidth]{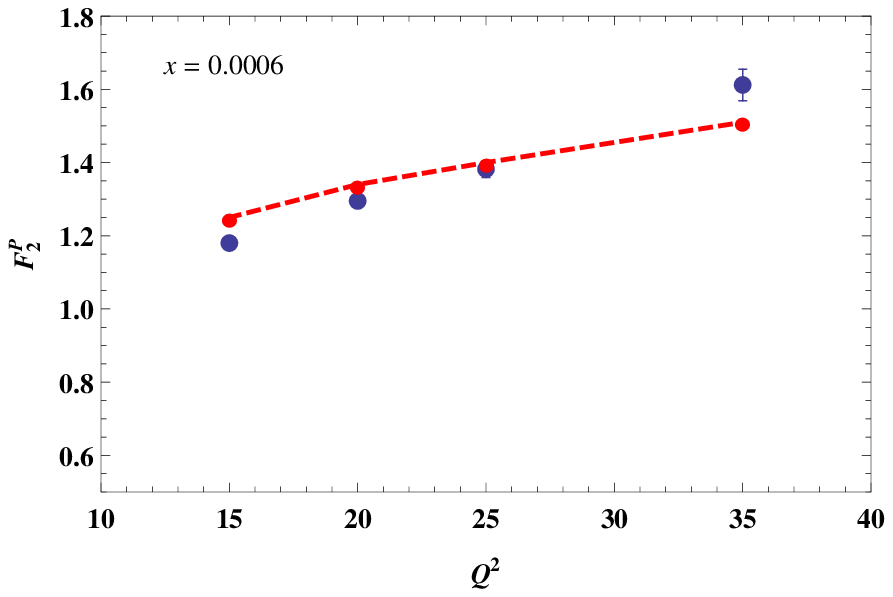}
\endminipage\hfill
\minipage{0.50\textwidth}
\includegraphics[width=\linewidth]{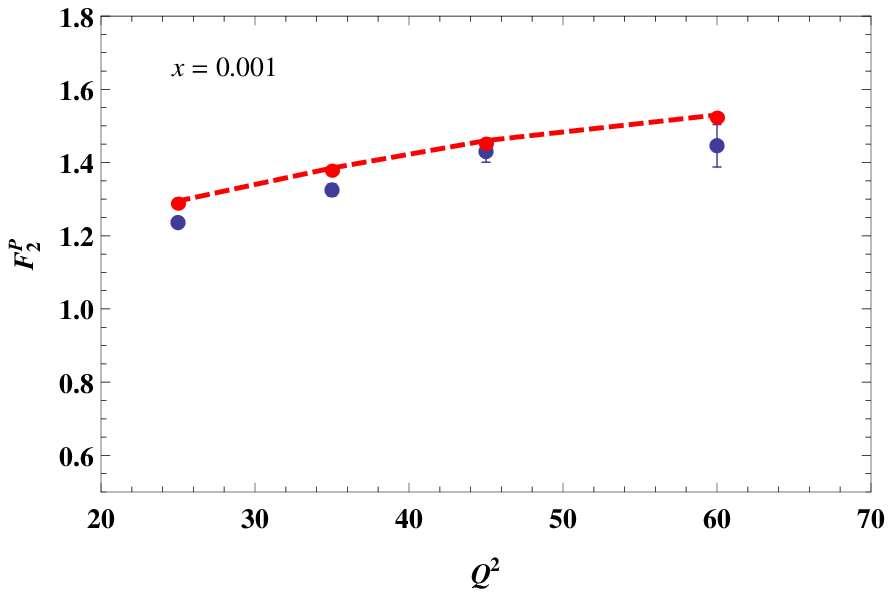}
\endminipage
\caption[Proton structure function $F_2^{P}(x,t)$]{Proton structure function $F_2^P{(x,t)}$ as a function of $Q^2$ for different fixed $x$ values by Lagrange's method with H1 data. Here the dash-connected line represents our analytical model.}
\end{figure}  
\begin{figure}[h]
\begin{center}
\includegraphics[width=4.5in]{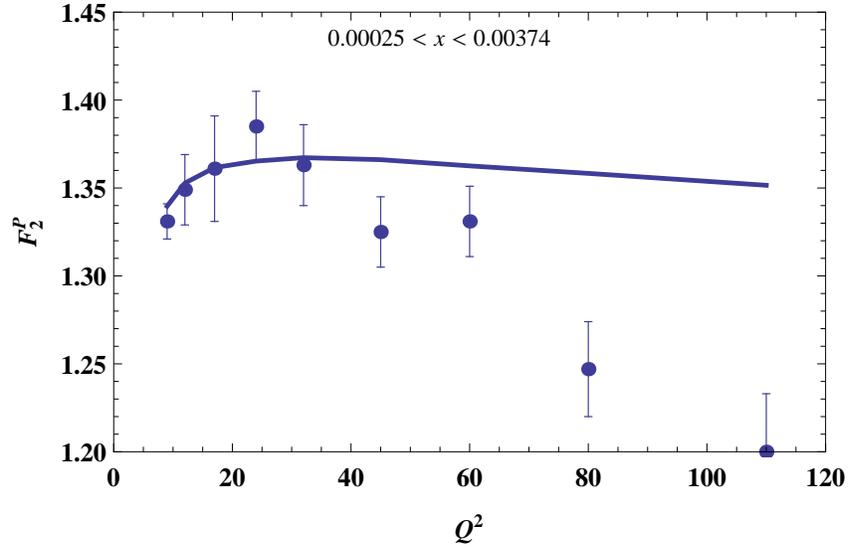}
\end{center}
\caption{Proton structure function$F_2^P(x,t)$ as function of $Q^2$ by Lagrange's method with ZEUS data. Experimental data is considered for different bins of $x$ and $Q^2$. Here the line represents our analytical model. }
\end{figure}

\begin{figure}[h]
\minipage{0.50\textwidth}
\includegraphics[width=\linewidth]{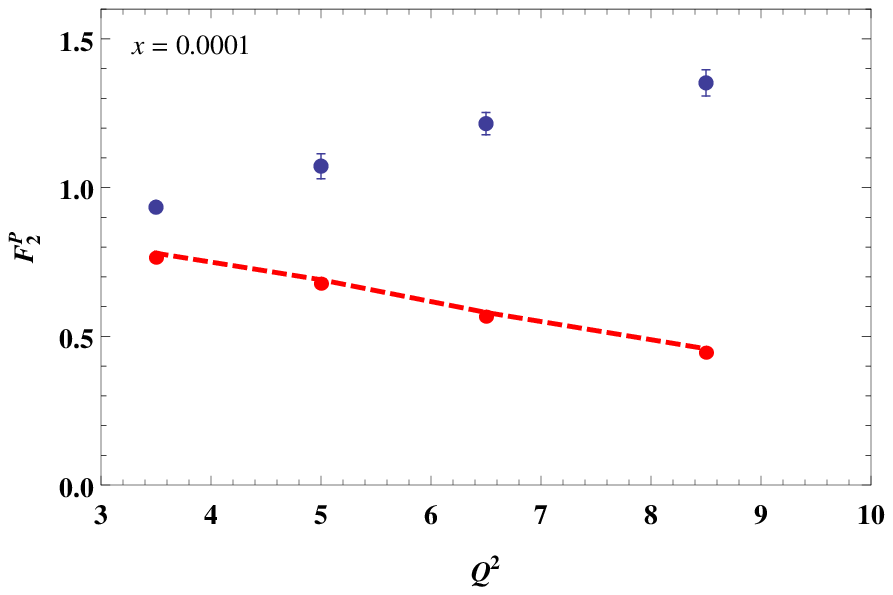}
\endminipage\hfill
\minipage{0.50\textwidth}
\includegraphics[width=\linewidth]{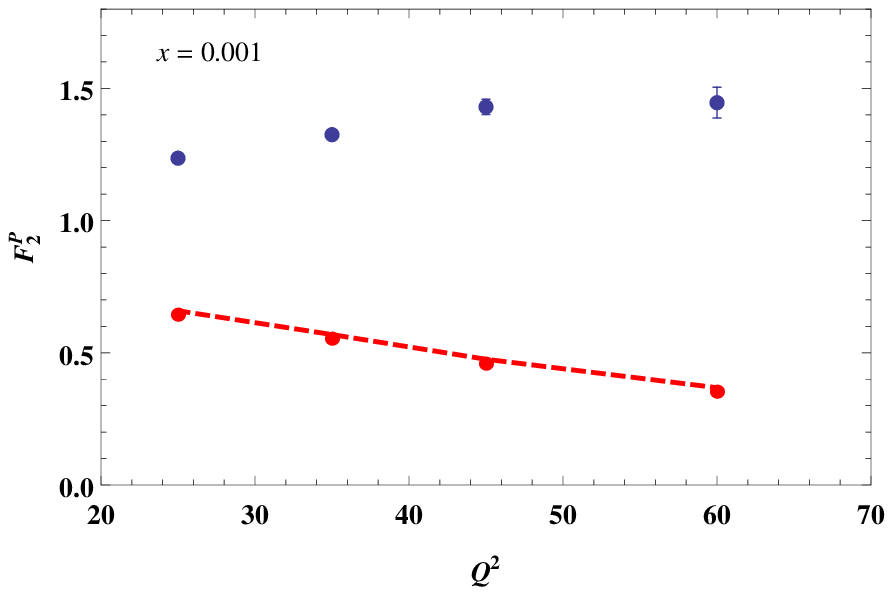}
\endminipage
\caption[Proton structure function $F_2^{P}(x,t)$]{Proton structure function $F_2^P{(x,t)}$ as a function of $Q^2$ for different fixed $x$ values by method of characteristics with H1 data. Here the dash-connected line represents our analytical model.}
\end{figure}  

\begin{figure}[b]
\minipage{0.50\textwidth}
\includegraphics[width=\linewidth]{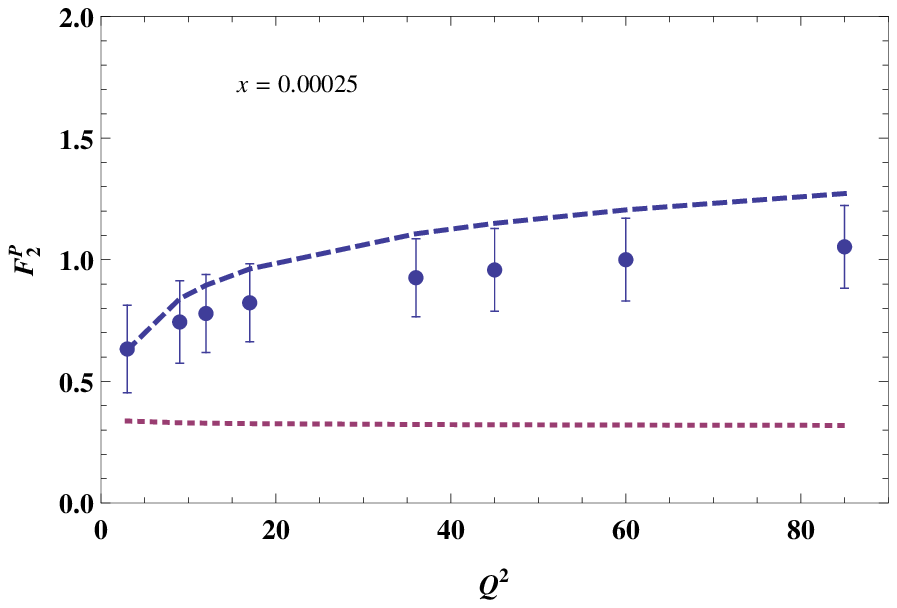}
\endminipage\hfill
\minipage{0.50\textwidth}
\includegraphics[width=\linewidth]{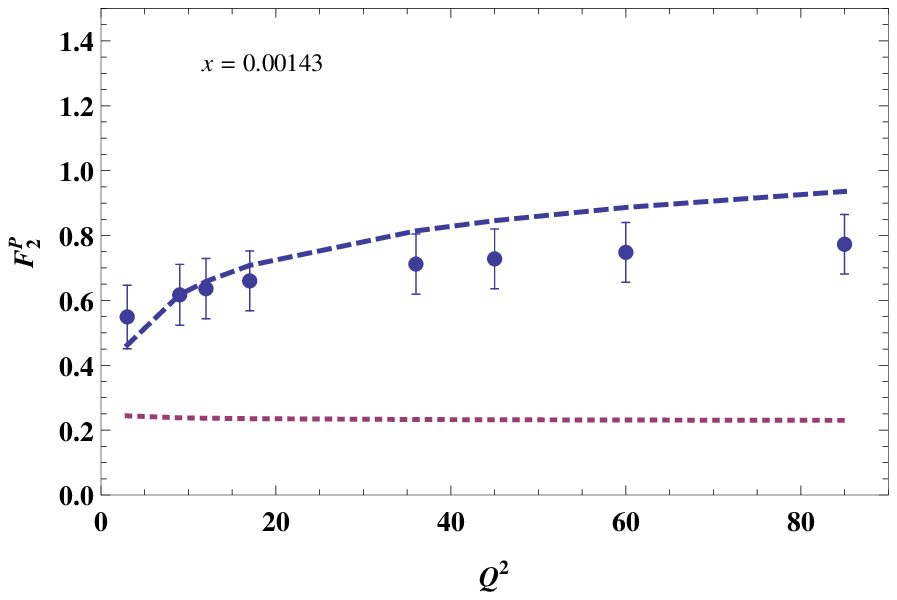}
\endminipage\hfill
\minipage{0.50\textwidth}
\includegraphics[width=\linewidth]{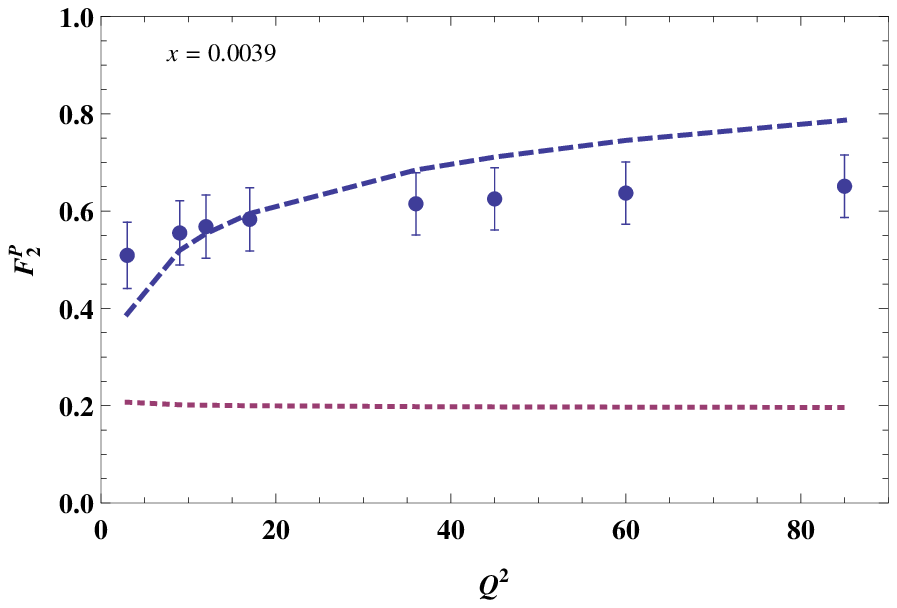}
\endminipage\hfill
\minipage{0.50\textwidth}
\includegraphics[width=\linewidth]{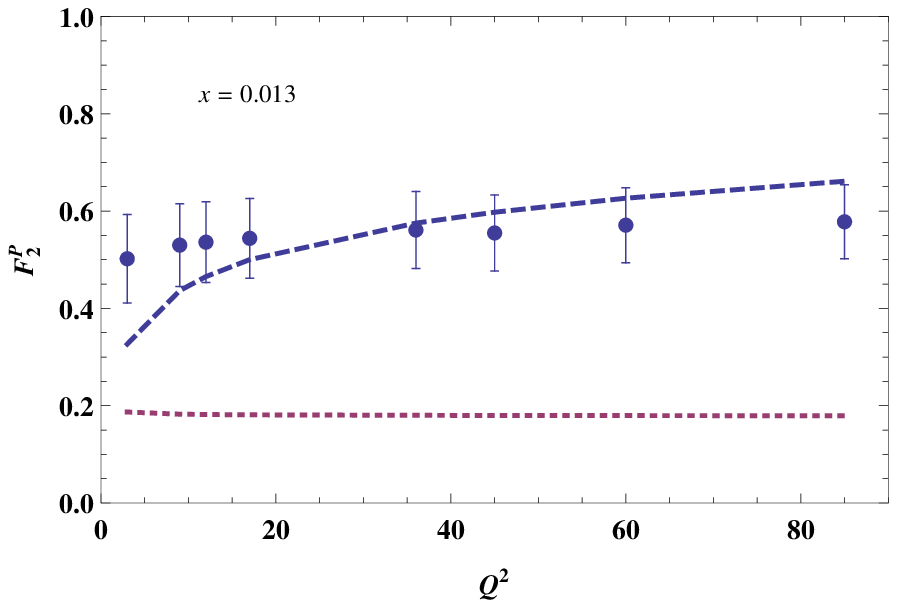}
\endminipage
\caption[Proton structure function $F_2^{P}(x,t)$]{Proton structure function $F_2^P{(x,t)}$ as a function of $Q^2$ for different fixed $x$ values by Lagrange's method and method of characteristics with NNPDF3.0 data with standard deviation. Here the dashed and dotted line represents our analytical model by Lagrange's method and method of characteristics respectively.}
\end{figure}


\end{document}